\documentclass[pre,aps,twocolumn]{revtex4}
\usepackage{epsfig}
\usepackage{bm}
\usepackage{color}
\usepackage[latin1]{inputenc}
\newcommand{\B}[1]{{\bm{#1}}}

\newcommand{\beq}{\begin{equation}}
\newcommand{\eeq}{\end{equation}}
\newcommand{\bea}{\begin{eqnarray}}
\newcommand{\eea}{\end{eqnarray}}

\begin{document}
\title{Direct Identification of the Glass Transition: Growing Length Scale and the Onset of
Plasticity}
\author{Einat Aharonov$^1$, Eran Bouchbinder$^2$, H. G. E. Hentschel$^3$, Valery Ilyin$^2$,  Nataliya Makedonska$^1$, Itamar Procaccia$^2$ and Nurith Schupper$^2$}
\affiliation{$^1$ The Department of Environmental Sciences and
Energy,\\ $^2$ The Department of Chemical Physics, The Weizmann
Institute of Science, Rehovot 76100, Israel,\\ $^3$ Dept. of
Physics, Emory University,  Atlanta, Georgia 30322.}
\begin{abstract}
 Understanding the mechanical properties of glasses remains elusive since the glass
 transition itself is not fully understood, even in well studied examples of glass formers
 in two dimensions. In this context we demonstrate here: (i) a direct evidence for
 a diverging length scale at the glass transition (ii) an identification of the glass
 transition with the disappearance of fluid-like regions and (iii) the appearance
 in the glass state of fluid-like regions when mechanical strain is applied.
 These fluid-like regions are associated with the onset of plasticity in the amorphous solid.
The relaxation times which diverge upon the approach to the glass
transition are related quantitatively.
\end{abstract}
 \maketitle

%\title{Direct Identification of the Glass Transition: Growing Length Scale and the Onset of
%Plasticity}
%\shorttitle{Glass Transition and Plasticity}
%\author{Einat Aharonov,\inst{1} Eran Bouchbinder, \inst{2 }H. G. E. Hentschel, \inst{3} Valery Ilyin, \inst{2}  Nataliya Makedonska, \inst{1} Itamar Procaccia \inst{2} and Nurith Schupper \inst{2}}
%\institute{\inst{1} Department of Environmental Sciences and Energy, Weizmann Institute of Science, Rehovot 76100, Israel\\
%\inst{2}The Department of Chemical Physics, The Weizmann
%Institute of Science, Rehovot 76100, Israel\\
%\inst{3} Dept. of Physics, Emory University,  Atlanta, Georgia 30322 }
%\pacs{61.43.Hv}{First pacs description}
%\pacs{05.45.Df}{Second pacs description}
%\pacs{05.70.Fh}{Third pacs description}
%
%
%\abstract{
% Understanding the mechanical properties of glasses remains elusive since the glass
% transition itself is not fully understood, even in well studied examples of glass formers
% in two dimensions. In this context we demonstrate here: (i) a direct evidence for
% a diverging length scale at the glass transition (ii) an identification of the glass
% transition with the disappearance of fluid-like regions and (iii) the appearance
% in the glass state of fluid-like regions when mechanical strain is applied.
% These fluid-like regions are associated with the onset of plasticity in the amorphous solid.
%The relaxation times which diverge upon the approach to the glass transition are related quantitatively
%to the diverging length scale.}
%\begin{document}
%\maketitle

{\bf Introduction:} The understanding of the mechanical properties
of amorphous solids cannot be dissociated from the elucidation of
their structure, including the so-called glass transition which has
eluded intensive research for quite some time. Plastic deformation,
for example, arises from a qualitatively different physics in
amorphous solids as compared to crystalline materials \cite{73KZ}.
Since dislocation glide is precluded by the lack of crystalline
order, much attention has been given to localized regions known as
Shear Transformation Zones (STZ) \cite{79Arg,98FL,06BLPa,06BLPb}
which are believed to be loci of plastic responses to increased
strain. In this Communication we demonstrate that a careful
identification of liquid-like structures sheds unexpected light on
the glass transition itself, including a sharp definition of the
transition, an identification of the long-sought diverging
length-scale \cite{05BBBCELLP} that accounts for the huge slowing
down associated with the transition, and finally, a connection
between the liquid-like defects and the onset of plastic
deformations in the glass state.

{\bf The system}: the glass former explored here is the well studied
system of a two-dimensional binary mixture of discs interacting via
a soft$1/r^{12}$ repulsion with a ``diameter" ratio of 1.4. This
two-dimensional model had been selected for simulation speed and,
more importantly, for the ease of interpretation. We refer the
reader to the extensive work done on this system \cite{99PH},
showing that it is a {\em bona fide} glass-forming liquid meeting
all the criteria of a glass transition. In short, the system
consists of an equimolar mixture of two types of particles with
diameter $\sigma_2=1.4$ and $\sigma_1=1$, respectively, but with the
same mass $m$. The three pairwise additive interactions are given by
the purely repulsive soft-core potentials
\begin{equation}
u_{ab} =\epsilon \left(\frac{\sigma_{ab}}{r}\right)^{12} \ , \quad a,b=1,2 \ , \label{potential}
\end{equation}
where $\sigma_{aa}=\sigma_a$ and $\sigma_{ab}=
(\sigma_a+\sigma_b)/2$. The cutoff radii of the interaction are set
at $4.5\sigma_{ab}$. The units of mass, length, time and temperature
are $m$, $\sigma_1$, $\tau=\sigma_1\sqrt{m/\epsilon}$ and
$T=\epsilon/k_B$, respectively, with $k_B$ being Boltzman's
constant. A total of $N=1024$ particles were enclosed in a square
box (of area $L^2$) with periodic boundary conditions, and our
simulations followed verbatim those described in \cite{99PH}, giving
us a most welcome check on the validity of our results. While Ref.
\cite{99PH} performed Molecular Dynamics, we ran both Molecular
Dynamics and Monte Carlo simulations. The results shown below are in
agreement between the two simulation methods. Ref. \cite{99PH} found
that for $T>0.5$ the system is liquid and  for lower temperatures
dynamical relaxation slows down. A precise glass transition had not
been identified in \cite{99PH}.
%%%%%%%%%%%%%%%%%%%%%%%%%%%
\begin{figure}
\centering
\epsfig{width=.35\textwidth,file=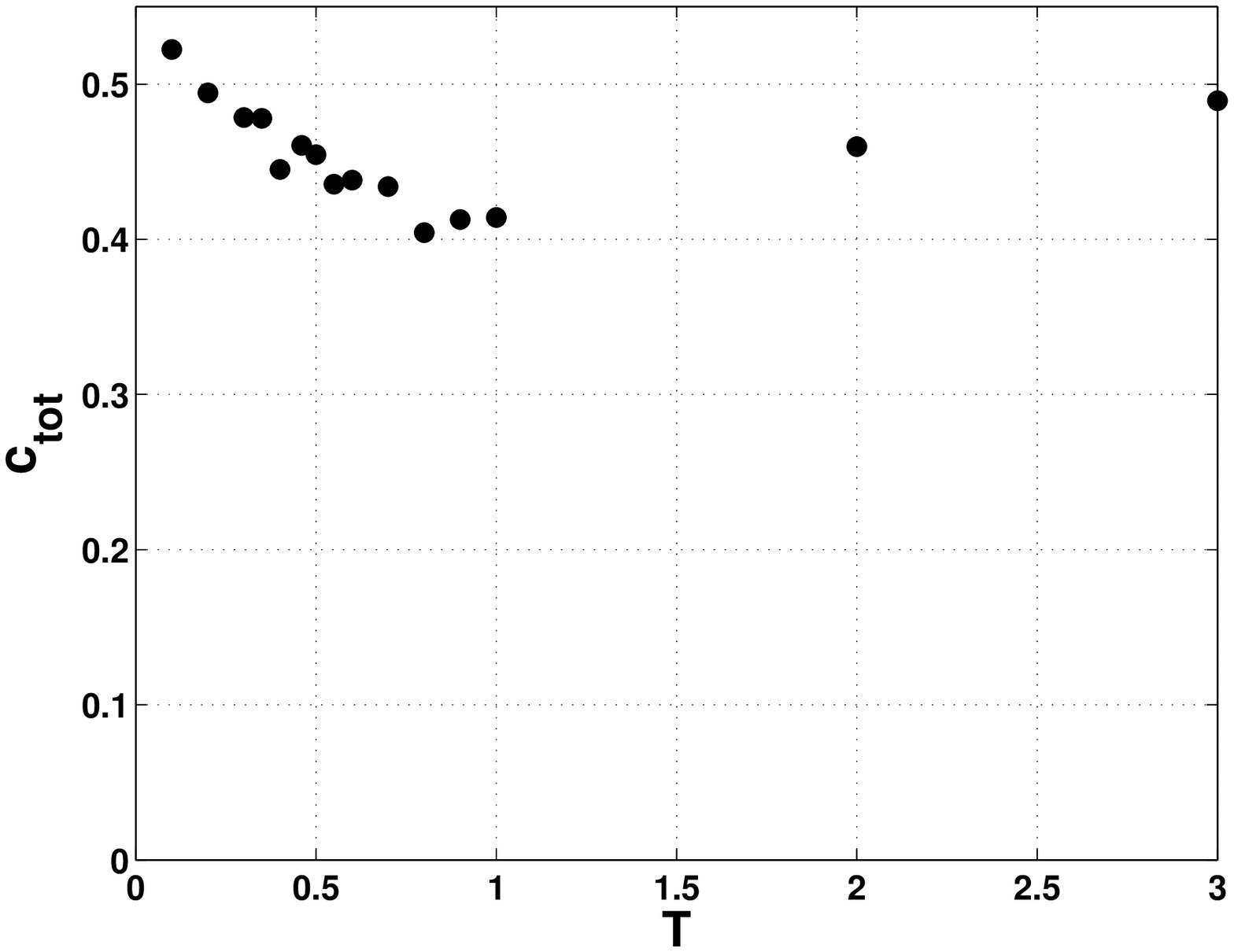}
\epsfig{width=.35\textwidth,file=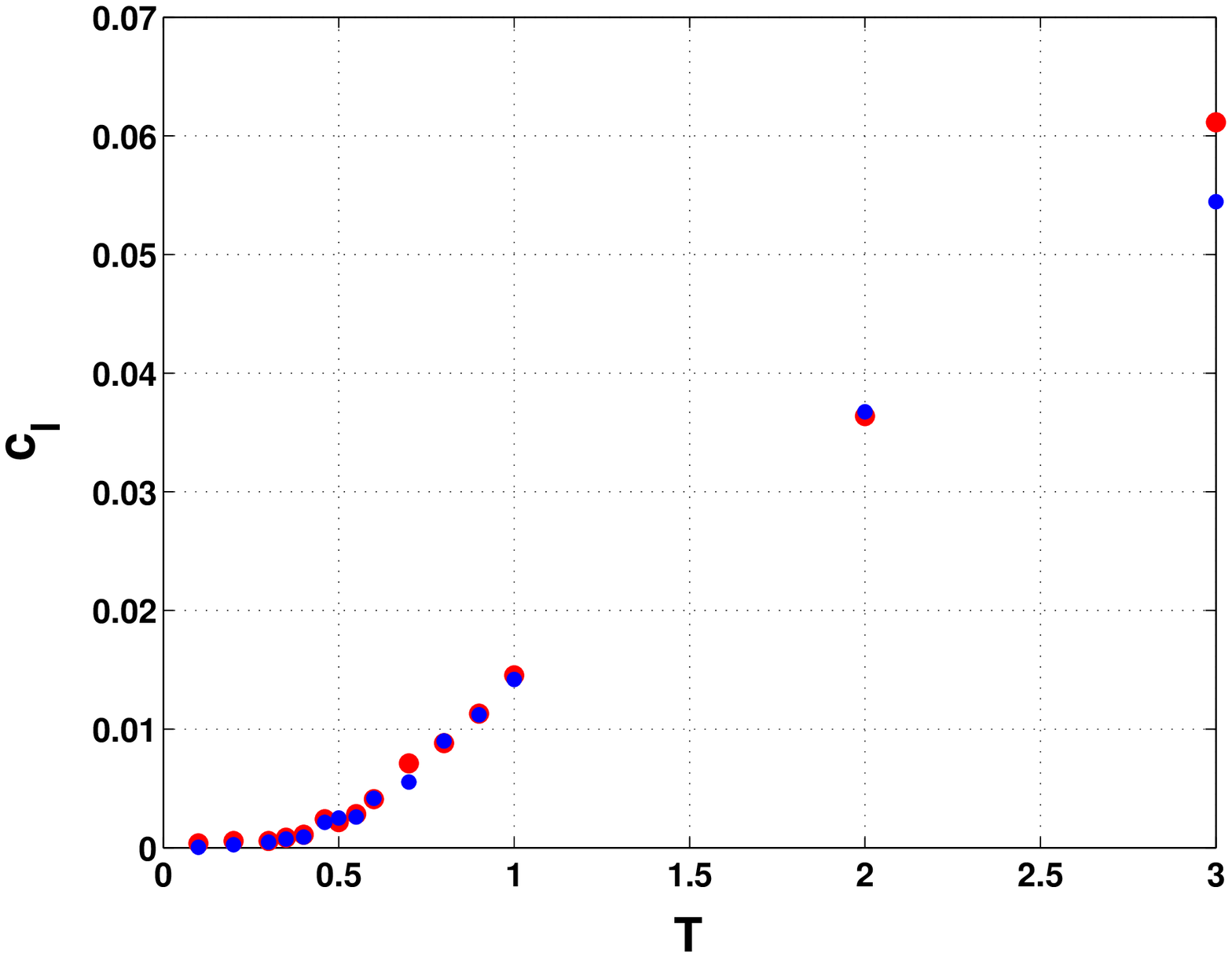}
\epsfig{width=.35\textwidth,file=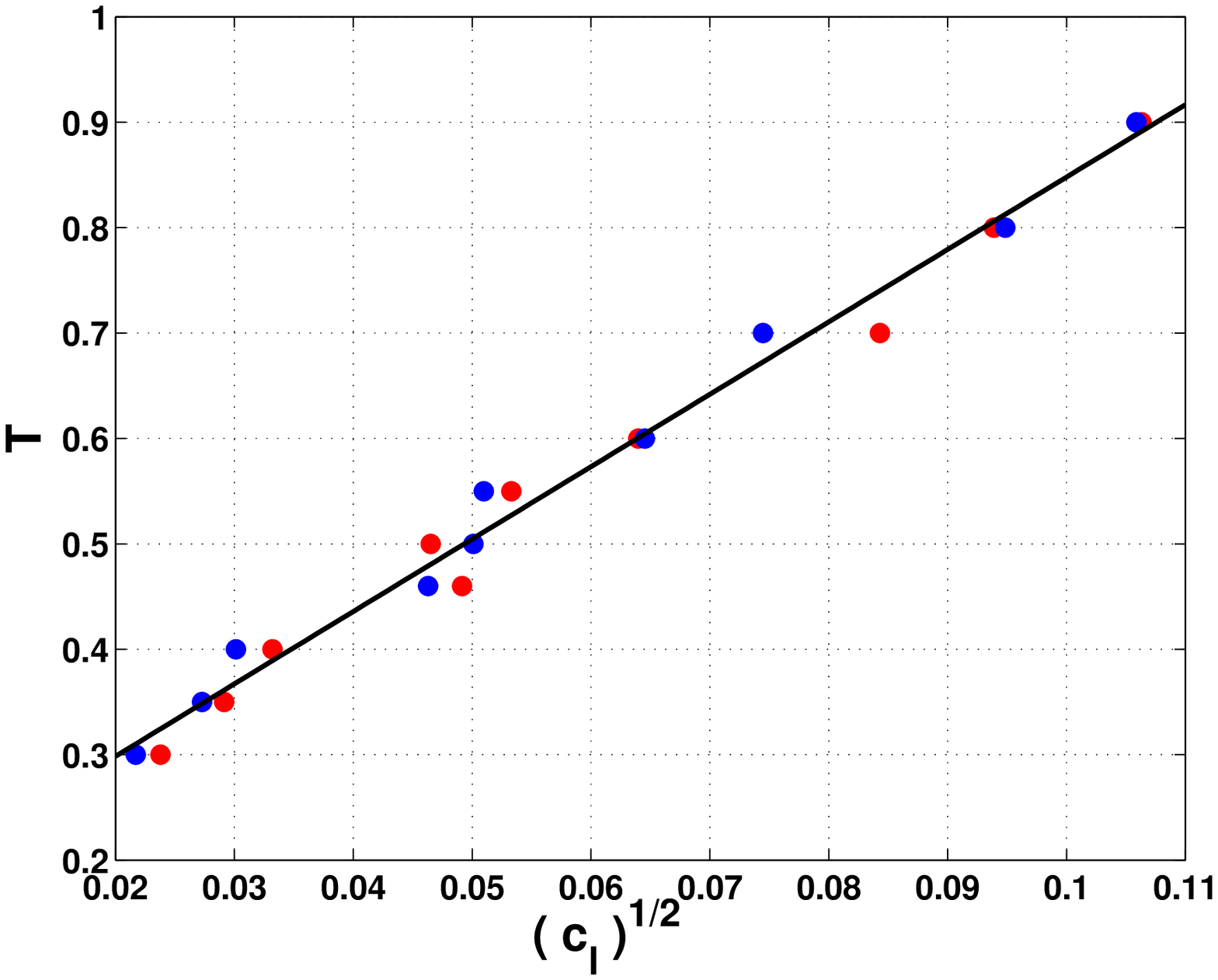}
\caption{Concentration of defects under slow cooling. Upper panel:
the average concentration of all the defects as a function of temperature in black dots. Middle panel: the concentration of the
liquid-like defects: large particles in
pentagons (red dots) and small particles in heptagons (blue dots).
Lower panel: the fit of the concentration of liquid-like defects as a function of temperature according
to Eq. (\ref{PLfit}).}
\label{concentrations}
\end{figure}
%%%%%%%%%%%%%%%%%%

\begin{figure}
\centering
\epsfig{width=.36\textwidth,file=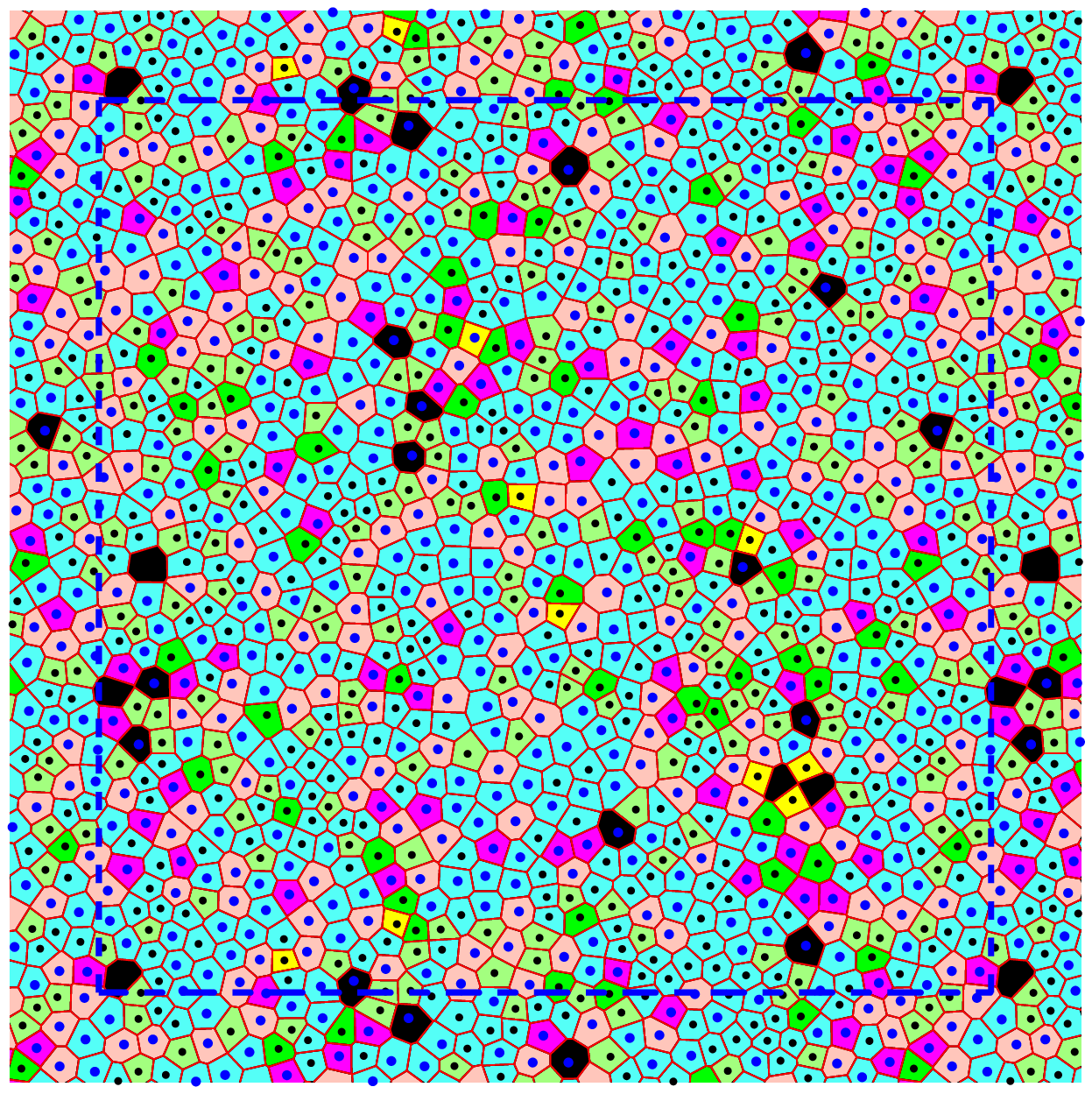}
\epsfig{width=.36\textwidth,file=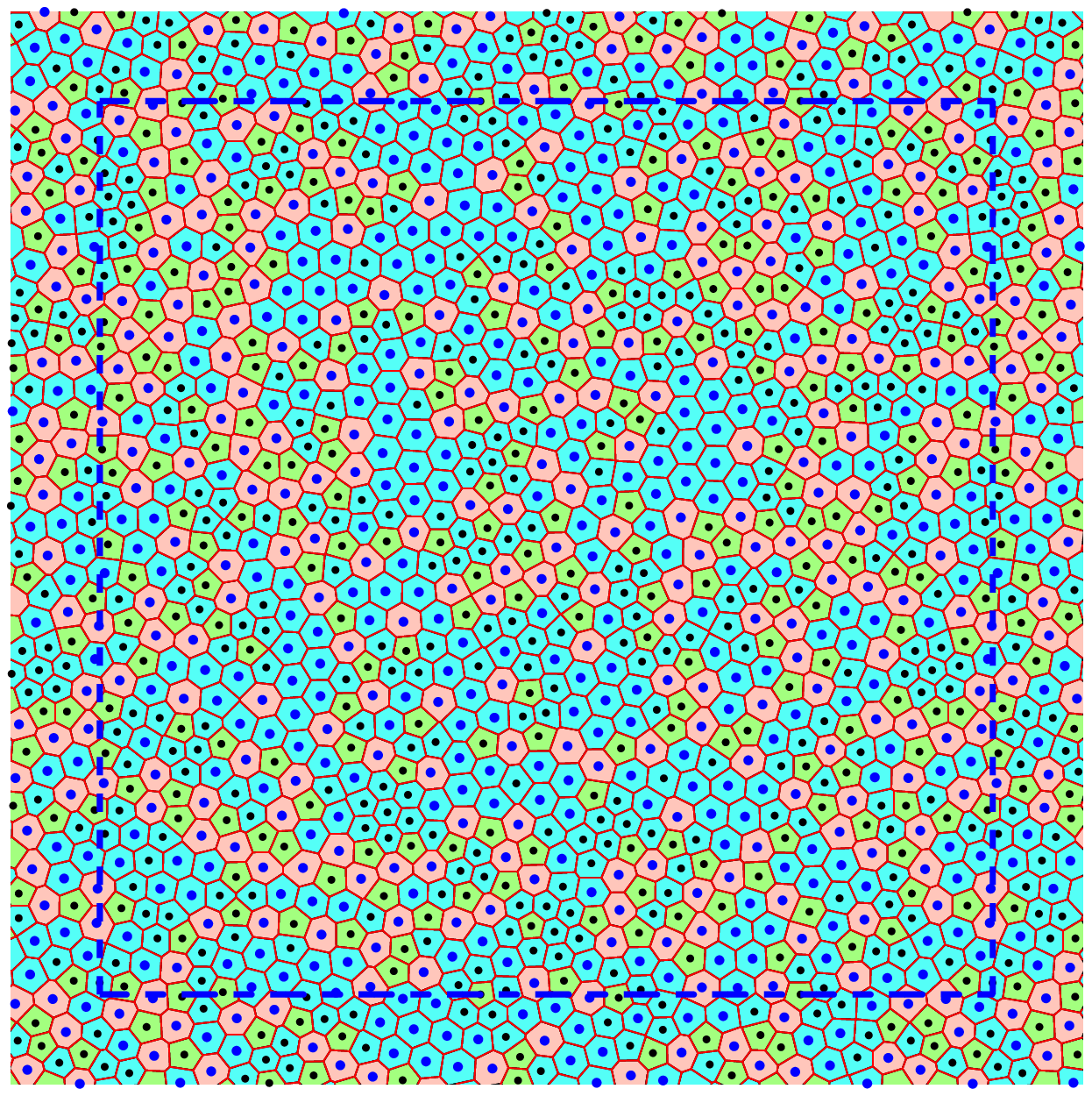}
\caption{Upper panel:  The Voronoi polygon construction in the liquid state at $T=3$  with the seven-color code used in this paper. Small particles in pentagons (heptagons) are light green (dark green) and large particles in pentagons (heptagons) are violet  (pink). Lower panel: a typical
Voronoi construction in the glass phase at $T=0.1$. Note the total disappearance of liquid-like
defects.}
\label{T=3}
\end{figure}

{\bf Analysis:} to visualize the structural features of this system
we employ the time-honored Voronoi polygon construction
\cite{89DAY}, where a polygon associated with any particle contains
all points closest to that particle than to any other particle. The
edges of such a polygon are the perpendicular bisectors of the
vectors joining the central particle. As had been noted in
\cite{89DAY,99PH}, the average coordination number is 6 at all
temperatures, and local coordination numbers other than 6 are
referred to as ``defects". Indeed, in previous work \cite{99PH} the
total concentration of  ``defects" was measured, see Fig.
\ref{concentrations} upper panel, and it was concluded that although
there is a slight decrease in this concentration when slowing down
sets in, nothing dramatic is observed. We propose here that a more
careful analysis is called for, in particular to distinguish between
``fluid-like" defects and ``glass-like" defects. We observe that
only in the liquid phase there exist {\bf small particles} enclosed
in heptagons (or even octagons), and {\bf large particles} enclosed
in pentagons (or even squares) (cf. Fig. \ref{T=3} upper panel). In
the glass phase we observe only defects of the opposite type, i.e.
small particles in pentagons and large particles in heptagons. We
propose that the concentration of the liquid-like defects is a
superior indicator of the glass transition in comparison with
relaxation times.
 {\bf The concentration $c_\ell$ of these liquid-like defects becomes so small in the glass phase that we cannot distinguish it from zero}
 (cf. Fig. \ref{T=3} lower panel), unless the glass is put under mechanical strain, as shown below.
 %%%%%%%%%%%%%%%%%%%%%%%%%%%
%%%%%%%%%%%%%%%%%%%%%%%%%%%%%
Associated with this concentration we can define a typical scale, $\xi$, according to
 \begin{equation}
 \xi \equiv 1/\sqrt{c_\ell} \ . \label{defxi}
 \end{equation}
Associated with the strong decrease in $c_\ell$ we observe a huge increase in the typical scale $\xi$,
in agreement with the tremendous slowing down of the dynamics.

In Fig.~ \ref{concentrations} middle panel we show $c_\ell$ as a function of the temperature for a protocol of slow cooling. For temperatures larger than 0.8 the concentration follows closely an exponential fit,
\begin{equation}
c_\ell = A \exp{-(\Delta E/T)} \ , \quad A\approx 0.094, ~\Delta E\approx 1.90 \ .  \label{expfit}
\end{equation}
For temperatures in the range $0.3<T<0.8$ we find an excellent fit to
\begin{equation}
c_\ell = B (T-T_g)^2\  , \quad  B\approx 0.02,~T_g=  0.16 \pm 0.02\ .\label{PLfit}
\end{equation}
The quality of this fit is demonstrated in  Fig. \ref{concentrations} lower panel. The fit (\ref{PLfit}) appears to identify a sharp glass transition $T_g=0.16\pm 0.02$; note, however, that there is no theoretical reason to expect that $c_\ell$ truly
vanishes at $T_g$, but it becomes so small that we indeed do not see a single liquid-like defect
in our finite-box simulations. We cannot exclude an exponentially small concentration that requires very much larger boxes to be observable.  We can however state that $T_g$ is finite since we demonstrate below that at temperatures lower than $T_g$ the system reacts to shear like a solid. With the same range of temperatures we fit
an apparently divergent length
\begin{equation}
\xi (T)\sim (T-T_g)^{-\nu} \ , \quad \nu=1 \ . \label{xi}
\end{equation}
The same caveat applies: in fact we can only state that $\xi(T)$ becomes exponentially larger than the system size.
%%%%%%%%%%%%%%%%%%%%%%%%%%%
\begin{figure}
\centering
\epsfig{width=.43\textwidth,file=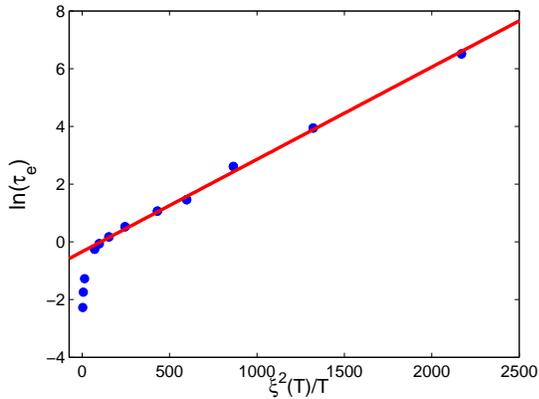}
\caption{The test of Eq. (\ref{taue}). The points are $\ln(\tau_e)$ as  a function of $\xi^2(T)/T$.
The solid line is a linear fit with $\tau_0 =0.71$ and $\Delta \mu =0.0055$.}
\label{tau}
\end{figure}
%%%%%%%%%%%%%%%%%%%%%%%%%%%

At this point we connect this new structural definition of the glass transition to the more commonly
considered criterion, i.e. the relaxation time (or viscosity), and show the one-to-one correspondence between the two. The relaxation times $\tau_e$ for the same system were measured in \cite{99aPH}via the decorrelation times of the scattering functions, as a function of  T, for $0.4<T<5$ . To connect between the relaxation time $\tau_e$ and and the length scale $\xi$ we assume as usual \cite{AG} that for the viscous fluid there exists a free energy of activation $\Delta G^*(T)$ associated with the relaxation event,
\begin{equation}
\tau_e = \tau_0 \exp (\Delta G^*(T)/T) \ ,
\end{equation}
where $\tau_0$ is a microscopic time scale of the order of a single particle vibration time. The free energy of activation is estimated as the number of Voronoi cells $N^*(T)$ involved in the relaxation event, times the (temperature independent) chemical potential per cell $\Delta \mu$, $\Delta G^*(T)\approx N^*(T) \Delta \mu$. The number $N^*$ is now
taken as the typical number of Voronoi cells in regions that are free of liquid like defects, and therefore
$N^*(T) \approx \pi \xi^2(T) /4 \bar {\Omega}$, where $\bar {\Omega}$ is the mean area of a Voronoi cell. We thus end up with the prediction
\begin{equation}
\tau_e = \tau_0 \exp (\pi \xi^2(T)\Delta \mu /4 \bar {\Omega} T) \ . \label{taue}
\end{equation}
Using now $\xi\approx 1/\sqrt{c_\ell}$ together with the fits (\ref{expfit}) and (\ref{PLfit}) we can
compare the prediction (\ref{taue}) to the measured relaxation times given in \cite{99aPH}. This comparison is presented in Fig. \ref{tau} , where we see an excellent agreement with $\tau_0 =0.71$ and $\Delta \mu=0.0055 $.
This value of $\tau_0$ is of the order of the expected particle vibration time \cite{99PH}.
At the highest values of the temperature, $T>2$, the concentration of liquid-like defects is too high, and
the law (\ref{taue}) does not apply.

Note that in our calculation we did not follow the traditional steps of Adam and Gibbs who, due to the lack of independent knowledge of $N^*$, assume that $N^*/s^*_c = N/S_c$ where $s^*_c$ is the configurational entropy of  the region involved in the relaxation, whereas $S_c$ is the total configurational entropy of the system. In our view this
step is questionable, since it appears that the configurational entropy per particle of the whole system, including the liquid-like defects, is higher than
the constrained configurational entropy per particle of the region which is free of liquid-like defects. Indeed, we predict that
the relaxation time diverges where $c_\ell\to 0$, (i.e. at $T_g$) whereas the Adam-Gibss formula predicts divergence only when $S_c\to 0$ (i.e. at the Kauzmann temperature $T_k$ if the latter exists \cite{48Kau}. We thus conclude that the two definitions of a glass transitions appear to agree on the existence of a finite value $T_g$ where the fluid becomes a glass.
We stress that it is considerably easier to determine $\xi(T)$ at low temperatures than to measure $\tau_e(T)$. The latter quantity becomes inaccessible to measurement due to jamming. The concentration of liquid-like defects is much easier to determine since the Voronoi tesselation remains highly mobile even in the glass phase. Minute changes in particle positions is reflected in changes in the Voronoi cells, allowing statistics to be accumulated at temperatures where the usual relaxation measurements become quite impossible. For this reason we are in a much better position to fit a finite value of $T_g$.

%%%%%%%%%%%%%%%%%%%%%%%%%%%
\begin{figure}
\centering
\epsfig{width=.39\textwidth,file=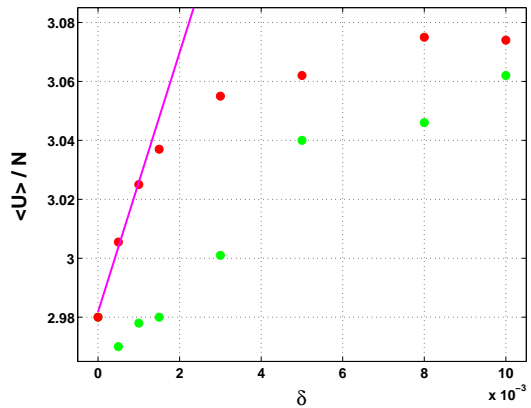}
\caption{The response of the system to shear. The measured energy of the system after the shear is applied is shown in red dots. The rhombic box is then returned
to a square box  and the energy at the end is shown in green dots.
For small values of $\delta$ the energy grows linearly in $\delta$. When the value of $\delta$ exceeds $10^{-3}$ the linear
dependence is lost, apparently due to a plastic response, and upon the return to the square box
memory is not lost and the energy does not return to the initial value. }
\label{response}
\end{figure}
%%%%%%%%%%%%%%%%%%%%%%%%%%
%%%%%%%%%%%%%%%%%%%%%%%%%%%
\begin{figure}
\centering
\epsfig{width=.38\textwidth,file=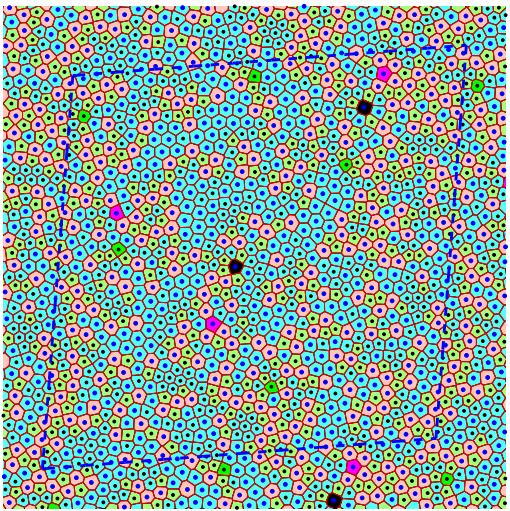}
\caption{A strained configuration after the linear elastic response is lost, $\delta =3\times 10^{-3}$. Notice the reappearance of liquid-like defects. }
\label{strained}
\end{figure}
%%%%%%%%%%%%%%%%%%%%%%%%%%%%%%%%%%%

{\bf The onset of plasticity:} having identified the glass transition with the disappearance of liquid-like defects, we proceed to strengthen this identification by analyzing the response of the glass state to a mechanical strain. In particular we show that for $T<T_g$ the system exhibits a finite shear
modulus as is expected from a solid. To this aim we applied strain on the system in the glass phase,
always starting from conditions as in Fig.~\ref{T=3} lower panel. The square box was strained to
a rhombic form keeping the area constant. This linear straining transformation is determined by the matrix $\B h$,
\begin{equation}
\B h\equiv \B 1 +\B \epsilon = \left(\begin{array}{cc}1+\delta&\sqrt{(1+\delta)^2-1}\\ \sqrt{(1+\delta)^2-1}&1+\delta \end{array}\right) \ ,
\end{equation}
where $\B \epsilon$ is the strain tensor. The linear elastic energy density associated with such a
transformation is
\begin{equation}
\frac{U_{\rm el}}{L^2} =\frac{1}{2} \lambda \epsilon^2_{kk} +\mu \epsilon_{ij}\epsilon_{ij} =(2\lambda+4\mu)\delta^2 +4\mu \delta \ , \label{Lame}
\end{equation}
where $\lambda$ and $\mu$ are the 2-dimensional Lam\'e coefficients.
In all our simulations
we applied the transformation with a chosen value of $\delta$, allowed the system to
relax by Monte Carlo steps in the strained configuration, and then we applied the inverse
transformation, returning to the square box. The results of these simulations are most revealing. For small values of $\delta$,  $\delta <  10^{-3}$, the system responds elastically, see Fig. \ref{response}.
In this regime the energy increases linearly with $\delta$ as expected from the leading
term in Eq. (\ref{Lame}).  As long as the response is elastic, it is also reversible: upon returning to the square box the energy returns back to its initial value.
On the other hand, when the value of $\delta$ exceeds $10^{-3}$ the linear
dependence is lost, but {\em not according to the elastic prediction} Eq. (\ref{Lame}) . Rather than
curving up as a function of $\delta$, the energy is curving down,  apparently due to a plastic response; Indeed, upon the return to the square box
memory is not lost and the energy does not return to the initial value. {\bf Most interestingly, for
values of strain where the linear elastic response is lost, the strained configuration exhibits
a significant concentration of liquid-like defects}, see Fig. \ref{strained}.
We propose that these liquid-like defects are the STZ that are the loci of plastic responses. Quite characteristically, once these appear, the mechanical response is not reversible, as seen
in Fig. \ref{response}. There is a remnant stress now, exhibiting the well known hysteretic behavior associated with the onset of plasticity.

{\bf Summary:} in summary, we have presented a novel view of the glass transition, identifying
the transition as associated with the disappearance of liquid-like defects.  A diverging correlation length was identified and measured.  We demonstrated that this divergence coincides with the divergence of the relaxation time as measured from standard time-correlation functions. The very same local arrangements of particles which is identified as liquid-like defects reappears once again
when plasticity sets in under strain.In a
subsequent publication we will present the statistical mechanics of this phenomenology,
explaining and substantiating further the findings presented here.

\acknowledgments

IP thanks Peter Harrowell for showing him the beauty of this subject. We benefitted from discussions and exchange of ideas with Michael Falk and Jim Langer. This work had been supported
in part by the Israel Science Foundation, by the Minerva Foundation and by the German-Israeli Foundation. EB is supported
by the Horowitz Center for Complexity Science.

\end{document}